\documentclass[10pt,number,sort&compress]{elsarticle}

\usepackage{comment}

\newcommand{\gsim}{{\;\raise0.3ex\hbox{$>$\kern-0.75em\raise-1.1ex\hbox{$\sim$}}\;}}
\newcommand{\lsim}{\hbox{ \raise3pt\hbox to 0pt{$<$}\raise-3pt\hbox{$\sim$} }}
\newcommand{\del}{\ifmmode{\nabla}               \else{$\nabla$ }               \fi}

\newcommand{\figdir}{.}

\usepackage{lmodern}

\usepackage{stfloats}

\usepackage{graphicx}

\usepackage[psamsfonts]{amssymb}

\usepackage{amssymb}

\usepackage{bm}

\usepackage{siunitx}

\journal{Nucl. Instr. and Meth. A}

\usepackage{url}
\usepackage{hyperref}
\usepackage{relsize}
\usepackage{bigints}

\newcommand{\vect}[1]{\boldsymbol{#1}}

\date{}

\usepackage{lineno}



\begin{document}

\begin{frontmatter}

\title{ A novel technique for the measurement of the avalanche fluctuations \\
  of a GEM stack using a gating foil}

\begin{frontmatter}
\title{Measurement of the electron transmission rate of the gating foil \\
for the TPC of the ILC experiment}
\end{comment}

%
%***************
%   Authors
%***************
%
\author[3]{M.~Kobayashi\corref{cor1}}
     \ead{makoto.kobayashi.exp@kek.jp}
     \cortext[cor1]{Corresponding author.
                           Tel.: +81 29 864 5376; fax: +81 29 864 2580.}
\author[7]{K.~Yumino}
%%%%%\author[7]{T.~Ogawa}
\author[7]{T.~Ogawa\fnref{20}}
\author[11]{A.~Shoji}
\author[7]{Y.~Aoki}
\author[1]{K.~Ikematsu\fnref{21}}
%%%%%\author[1]{K.~Ikematsu}
%%%%%\author[1,13]{K.~Ikematsu}
\author[1]{P.~Gros\fnref{22}}
\author[8]{T.~Kawaguchi}
\author[12]{D.~Arai}
\author[12]{M.~Iwamura}
\author[12]{K.~Katsuki}
\author[12]{A.~Koto}
\author[12]{M.~Yoshikai}
\author[3]{K.~Fujii}
\author[1]{T.~Fusayasu}
\author[4]{Y.~Kato}
\author[10]{S.~Kawada\fnref{21}}
\author[3]{T.~Matsuda}
\author[7]{T.~Mizuno\fnref{21}}
\author[7]{J.~Nakajima}
\author[11]{S.~Narita}
\author[11]{K.~Negishi}
\author[17]{H.~Qi}
\author[5]{R.D.~Settles}
\author[1]{A.~Sugiyama}
\author[8]{T.~Takahashi}
\author[3]{J.~Tian\fnref{25}}
\author[6]{T.~Watanabe}
\author[9]{R.~Yonamine\fnref{21}}
%
%***************
%   Addresses
%***************
%
\address[3]{High Energy Accelerator Research Organization (KEK), Tsukuba 305-0801, Japan}
\address[7]{The Graduate University for Advanced Studies (Sokendai), Tsukuba 305-0801, Japan}
\address[11]{Iwate University, Iwate 020-8551, Japan}
\address[1]{Saga University, Saga 840-8502, Japan}
\address[8]{Hiroshima University, Higashi-Hiroshima 739-8530, Japan}
\address[12]{Fujikura Ltd., 1440, Mutsuzaki, Sakura-city, Chiba 285-8550, Japan}
\address[4]{Kindai University, Higashi-Osaka 577-8502, Japan}
\address[10]{Deutsches Elektronen-Synchrotron (DESY), D-22607 Hamburg, Germany}
\address[17]{Institute of High Energy Physics, Chinese Academy of Sciences,
Beijing 100049, China}
\address[5]{Max Planck Institute for Physics, DE-80805 Munich, Germany}
\address[6]{Kogakuin University, Shinjuku 163-8677, Japan}
\address[9]{Department of Physics, Tohoku University, Sendai 980-8578, Japan}

\fntext[20]{Now at High Energy Accelerator Research Organization (KEK), Tokai 319-1106, Japan}
\fntext[21]{Now at High Energy Accelerator Research Organization (KEK), Tsukuba 305-0801, Japan}
%%%%%\fntext[21]{Now at Institute of Multidisciplinary Research for Advanced Materials (IMRAM), Tohoku University, Sendai 980-8577, Japan.}
%%%%%\address[13]{now at Institute of Multidisciplinary Research for Advanced Materials (IMRAM), Tohoku University, Sendai 980-8577, Japan}
\fntext[22]{Now at Department of Physics, Engineering Physics \& Astronomy,
                    Queen's University, Kingston, Ontario K7L 3N6, Canada.}
%%%%%\fntext[23]{Now at High Energy Accelerator Research Organization (KEK), Tsukuba 305-0801, Japan}
%%%%%\fntext[24]{Now at High Energy Accelerator Research Organization (KEK), Tsukuba 305-0801, Japan}
\fntext[25]{Now at ICEPP, University of Tokyo, Hongo, Tokyo 113-0033, Japan.}
where $\vect{x} \equiv (x,y)$, with the z-coordinate axis along the laser beam,
$t$ denotes time in the duration of a pulse measured from its start time ($t^*$).
The discrete variable $t^*$ is the time stamp for each pulse and the  
symbol $\left< \cdots \right>$ for a function of $t^*$ in the following denotes the average over pulses.

First, we neglect its $z$-dependence in a small interval $\Delta z$:
\begin{equation}
  \phi(\vect{x},z,t;t^*) = \phi(\vect{x},t;t^*) 
\end{equation}
at any given time ($t$) during a pulse.
Next, we suppose that the time dependence of photon flux is independent of $\vect{x}$:
\begin{equation}
  \phi(\vect{x},t;t^*) = f(\vect{x};t^*) \cdot a(t) \;  
\end{equation}
with
\begin{equation}
  \int_0^T a(t) \; dt = 1 \;
\end{equation}
where $T$ denotes the duration of pulses.
It should be noted that the function $a(t)$ is bell-shaped with a peak near the middle of the pulse duration, and with a small positive skewness
in our case. 
Its duration is about 3-4 nsec at FWHM, which corresponds to the width in space (beam length) of $\sim$1 m (FWHM)
along z-axis.
Therefore, the z-dependence of the flux at any given $t$ is small
and Eq.~(D.1) is expected to be a reasonable approximation 
since the beam length is large compared to $\Delta z$ ($\sim$10 mm, in our case).

Finally,
the shape of photon density in the $x$-$y$ plane is assumed to be identical for all the pulses:
\begin{equation}
f(\vect{x};t^*) = f_0(\vect{x}) \cdot \left( 1 + \epsilon(t^*) \right)
\end{equation}
where $f_0(\vect{x})$ is the local fluence at $\vect{x}$ averaged over pulses with $\left< \epsilon(t^*) \right>$ = 0, and
\begin{comment}
The local fluence (number of photons/unit area/pulse) is then given by
\begin{eqnarray}
  f(\vect{x};t^*) &\equiv& \int \phi(\vect{x},t;t^*) \; dt\\
  &=& f_0(\vect{x}) \cdot \left( 1 + \epsilon(t^*) \right) 
\end{eqnarray}
\end{comment}
%%%with
\begin{eqnarray}
  \epsilon(t^*) &=& \frac{f(\vect{x};t^*) - f_0(\vect{x})}{f_0(\vect{x})} \\
  &\equiv& \frac{\delta f(\vect{x};t^*)}{f_0(\vect{x})} \;.
\end{eqnarray}

In short, we consider repeated irradiations on a thin layer of impurity molecules by a series of laser pulses 
with an identical shape of the beam profile, and with each being characterized only by the total number of photons, 
except for the statistical fluctuations in the number of photons distributed to each section in the profile.

Now, let us consider the probability that an impurity molecule staying at $\vect{x}$ is ionized in a time interval [$t_2, \; t_2 + \Delta t$]  
measured from the beginning of a pulse ($t^*$).
We define $\sigma_{\rm ex}$ to be the cross section of the impurity molecule for the transition from the ground state to the excited state,
and $\sigma_{\rm i}$ to be that for the transition from the excited state to the ionized (continuum) state.
Since the molecule needs to have been excited before $t_2$ the probability is given by
\begin{eqnarray}
  P(t_2) \; \cdot \Delta t &=& \Delta t \cdot f(\vect{x}; t^*) \cdot a(t_2) \cdot \sigma_{\rm i} \cdot
  \int_0^{t_2} f(\vect{x}; t^*) \cdot a(t_1) \cdot \sigma_{\rm ex} \; dt_1 \\
  &=& \Delta t \cdot f^2(\vect{x}; t^*) \cdot \sigma_{\rm i} \cdot \sigma_{\rm ex} \cdot a(t_2) \cdot
  \int_0^{t_2} a(t_1) \; dt_1 \;. 
  \end{eqnarray}
%%%%%%%The probability for the molecule to be ionized during the pulse is given by, with $T$ denoting the time at the end of the pulse,

The probability for the molecule to be ionized during the pulse is then given by
\begin{eqnarray}
  \int_0^T P(t_2) \; dt_2 &=& \sigma_{\rm i} \cdot \sigma_{\rm ex} \cdot f^2(\vect{x}; t^*) \cdot
                 \int_0^T a(t_2) \; dt_2 \; \int_0^{t_2} a(t_1) \; dt_1 \\
   &=& \sigma_{\rm i} \cdot \sigma_{\rm ex} \cdot f^2(\vect{x}; t^*) \cdot
                 \int_0^T \frac{d}{dt_2} \Biggl( \int_0^{t_2} a(t^{\prime}) \; dt^{\prime} \Biggr) \; dt_2 \cdot \int_0^{t_2} a(t_1) \; dt_1 \\
   &=& \sigma_{\rm i} \cdot \sigma_{\rm ex} \cdot f^2(\vect{x}; t^*) \cdot
                 \int_0^T A(t_2) \cdot \frac{dA(t_2)}{dt_2} \; dt_2 
\end{eqnarray}
with
\begin{equation}
  A(t_2) \equiv \int_0^{t_2} a(t^{\prime}) \; dt^{\prime} \;.
\end{equation}
The integral can be evaluated by partial integration:
\begin{eqnarray}
  \int_0^T A(t_2) \cdot \frac{dA(t_2)}{dt_2} \; dt_2  &=& \Bigl[ A^2(t_2) \Bigr]_0^T - \int_0^T A(t_2) \cdot \frac{dA(t_2)}{dt_2} \; dt_2 \\
  &=& \frac{1}{2} \cdot \Bigl[ A^2(t_2) \Bigr]_0^T \\
  &=& \frac{1}{2} \cdot \Biggl( \int_0^T a(t^\prime) \; dt^\prime \Biggr)^2 \\ 
  &=& \frac{1}{2} \;. 
\end{eqnarray}
Multiplying the number of target molecules, the average number of ionizations around $\vect{x}$ is given by
\begin{equation}
  \overline{n}(\vect{x}; t^*) \; d\vect{x} = c \cdot f^2(\vect{x}; t^*) \; d\vect{x} \;
\end{equation}
where
\begin{equation}
  c = \frac{1}{2} \cdot \sigma_{\rm i} \cdot \sigma_{\rm ex} \cdot \rho \cdot \Delta z 
\end{equation}
with  $\rho$, the density of the impurity molecules at the ground state.
%%%%%We neglect in what follows the statistical fluctuations of $n(\vect{x}, t^*)$ around $\overline{n}(\vect{x}, t^*)$.

The average total number of created electrons for the pulse with a time stamp $t^*$ is then given by
\begin{eqnarray}
  \overline{n}(t^*) &=& \int \overline{n}(\vect{x};t^*) \; d\vect{x} \\
  &=& c \cdot \int f^2(\vect{x}; t^*) \; d\vect{x} \\
  &=& c \cdot \bigl( 1 + \epsilon(t^*) \bigr)^2 \cdot \int f_0^2(\vect{x}) \; d\vect{x}
\end{eqnarray}
and its average over $t^*$ (pulses) is 
\begin{eqnarray}
  \left<\left< n \right>\right> &\equiv& \left< \overline{n}(t^*) \right>  \\
  &=& c \cdot \left< \bigl( 1 + \epsilon(t^*) \bigr)^2 \right> \cdot \int f_0^2(\vect{x}) \; d\vect{x} \\
  &=& c \cdot \left( 1 + \sigma^2 \right) \cdot \int f_0^2(\vect{x}) \; d\vect{x} \; 
\end{eqnarray}
with
\begin{equation}
  \sigma^2 \equiv \left< \epsilon^2(t^*) \right> \;. 
\end{equation}

Actually, $\sigma$ is the standard deviation of the fractional variation of the
total number of photons in a pulse:
\begin{equation}
  \sigma  =  \frac{\sqrt{\left< \bigl( \delta f(\vect{x};t^*) \bigr)^2 \right>}}{f_0(\vect{x})}
  \equiv \frac{\sigma_{\rm f}(\vect{x})}{f_0(\vect{x})} = \frac{\sigma_{\rm F}}{\left< F \right>} \;
\end{equation}
where
\begin{equation}
  \left< F \right> \equiv \int f_0 (\vect{x}) \; d\vect{x} 
\end{equation}
and 
\begin{eqnarray}
  \sigma_{\rm F}^2 &=& \left< \bigl( \delta F(t^*) \bigr)^2 \right> \\
  &\equiv& \left< \bigl( F(t^*) - \left< F \right> \bigr)^2 \right> \\
  &=& \left< \epsilon^2(t^*) \right> \cdot \left< F \right>^2 \\  
  &=& \sigma^2 \cdot \left< F \right>^2 
\end{eqnarray}
with
\begin{eqnarray}
F(t^*) &\equiv& \int f(\vect{x};t^*) \; d\vect{x} \\
  &=& \int \bigl( 1 + \epsilon(t^*) \bigr) \cdot f_0(\vect{x}) \; d\vect{x} \\
  &=& \bigl( 1 + \epsilon(t^*) \bigr) \cdot \left< F \right> \;. 
\end{eqnarray} 
Here
 $\left< F \right>$ represents the average total number of photons in a pulse, which is proportional to the average laser pulse energy.
It should be noted that the average number of ionizations is proportional to the square of the average pulse energy
since $\left<\left< n \right>\right>$ is quadrupled while $\left< F \right>$ is doubled when $f_0(\vect{x})$ is uniformly doubled over $\vect{x}$, for example,
according to Eqs.~(D.24) and (D.27).

The distribution of $\epsilon(t^*)$ is more or less Gaussian
for our new laser system of a similar type (Litron Nano S 120-20-FHG)\footnote{
  Courtesy of Litron Lasers Ltd., Rugby, United Kingdom.
  }
although we did not have a chance to measure it for New Wave Research Polaris II used in the present experiment.
Let us therefore suppose further that the random variable
\begin{equation}
  r \equiv \epsilon(t^*)
     = \frac{\delta F(t^*)}{\left< F \right>} 
\end{equation}
is Gaussian distributed with a standard deviation of $\sigma$\footnote{
  In fact, the value of $r = \epsilon(t^*)$ is discrete since $F(t^*)$ is an integer (total number of photons in a pulse).
  We assume the random variable $r$ to be continuous because of the large number of $\left< F \right>$ ($\sim 10^{13}$).
}.
\begin{comment}
\begin{equation}
  P(r) = \frac{1}{\sqrt{2 \pi} \cdot \sigma} \cdot \exp \left( - \frac{r^2}{2\sigma^2} \right) \;.
\end{equation}
\end{comment}
The assumption simplifies the following calculations since
\begin{equation}
  \left< \epsilon^6(t^*) \right> = 15 \cdot \sigma^6 , \;\;
  \left< \epsilon^4(t^*) \right> = 3 \cdot \sigma^4 , \;\;
  \left< \epsilon^2(t^*) \right> \equiv \sigma^2
\end{equation}
while
\begin{eqnarray}
  \left< \epsilon^5(t^*) \right> = \left< \epsilon^3(t^*) \right> = \left< \epsilon(t^*) \right> = 0 
\end{eqnarray}
with $\sigma = \sigma_{\rm F} / \left< F \right>$.

The variance of the average number of created electrons in this case is given by
\begin{eqnarray}
  \left< \bigl( \overline{n}(t^*) - \left<\left< n \right>\right> \bigr)^2 \right> &=& c^2 \cdot
  \left< \left( \left(1 + \epsilon(t^*) \right)^2 - \left( 1+\sigma^2 \right) \right)^2 \right> \cdot
  \left( \int f_0^2(\vect{x}) \; d\vect{x} \right)^2 \\
  &=& 4 \cdot c^2 \cdot \sigma^2 \cdot \left( 1 + \frac{1}{2} \cdot \sigma^2 \right) \cdot \left( \int f_0^2(\vect{x}) \; d\vect{x} \right)^2 \;
\end{eqnarray}
and the relative variance is, from Eqs.~(D.24) and (D.39), 
\begin{eqnarray}
  \frac{\left< \bigl( \overline{n}(t^*) - \left<\left< n \right>\right> \bigr)^2 \right>}{\left<\left< n \right>\right>^2}
  &=& 4 \cdot \sigma^2 \cdot \frac{1 + \frac{1}{2} \cdot \sigma^2}{\left( 1 + \sigma^2 \right)^2} \\
  &=& 4 \cdot \frac{\sigma_{\rm F}^2}{\left< F \right>^2} \cdot \frac{1 + \frac{1}{2} \cdot \frac{\sigma_{\rm F}^2}{\left< F \right>^2}}
       {\left( 1 + \frac{\sigma_{\rm F}^2}{\left< F \right>^2} \right)^2} \;.
\end{eqnarray}
Similarly,
\begin{eqnarray}
  \left< \bigl( \overline{n}(t^*) - \left<\left< n \right>\right> \bigr)^3 \right> &=& c^3 \cdot
  \left< \left( \bigl(1 + \epsilon(t^*) \bigr)^2 - \left( 1+\sigma^2 \right) \right)^3 \right> \cdot
  \left( \int f_0^2(\vect{x}) \; d\vect{x} \right)^3 \\
&=&  24 \cdot c^3 \cdot \sigma^4 \cdot \left( 1 + \frac{1}{3} \cdot \sigma^2 \right) \cdot \left( \int f_0^2(\vect{x}) \; d\vect{x} \right)^3 \;
\end{eqnarray}
and the skewness is given by, from Eqs.~(D.39) and (D.43),  
\begin{eqnarray}
%%  \frac{\left< \left( Q(t^*) - \left< Q \right> \right)^3 \right>}{\left< \left( Q(t^*) - \left< Q \right> \right)^2 \right>^{\frac{3}{2}}}
  \frac{\left< \bigl( \overline{n}(t^*) - \left<\left< n \right>\right> \bigr)^3 \right>}
       {\left< \bigl( \overline{n}(t^*) - \left<\left< n \right>\right> \bigr)^2 \right>^{\frac{3}{2}}}
   &=& 3 \cdot \sigma \cdot \frac{1+\frac{1}{3} \cdot \sigma^2}{\left( 1+\frac{1}{2} \cdot \sigma^2 \right)^{\frac{3}{2}}} \\
   &=& 3 \cdot \frac{\sigma_{\rm F}}{\left< F \right>} \cdot \frac{1+\frac{1}{3} \cdot \frac{\sigma_{\rm F}^2}{\left< F \right>^2}}
       {\left( 1+\frac{1}{2} \cdot \frac{\sigma_{\rm F}^2}{\left< F \right>^2} \right)^{\frac{3}{2}}} \;.
\end{eqnarray}

The specification for the pulse energy stability of the laser used (New Wave Research Polaris II)
is $\pm$ 6\% for 98\% of pulses.
If the distribution of $r = \delta F(t^*) / \left< F \right>$ is Gaussian, the 98\%-containment corresponds to $\pm$ 2.3 standard deviations
and the limit of $\pm$ 6\% gives 2.6\% for $\sigma_{\rm F}/\left< F \right>$.
The relative variance and the skewness of the $\overline{n}$ distribution are, therefore,
expected to be about 0.0027 and 0.078, respectively, from Eqs.~(D.41) and (D.45).
Our model, though possibly oversimplified, gives the values close to the observed ones:
0.0033 for the relative variance $C^2$ (see Eqs.~(8) and (9), and Fig.~5) and $\sim$0.1 for the skewness $S_{\left< n\right>}$,
that is expected to be the asymptotic value of the skewness of the signal charge distribution ($S$)
for large $\left< Q \right>$ (see Fig.~7 and text).

The same results can be obtained easily
by assuming a simple structure of the pulsed laser beam, i.e. 
a rectangular distribution both in space and time.
This simplification is justified because the laser beam profile in the $x$-$y$ plane at any given time during a pulse can be
assumed to be a set of independent square column-like distributions like a lego plot as far as the impurity molecules are {\it stationary\/} targets,
and the number of local ionizations $\overline{n}(\vect{x}, t^*)$ does not depend on the shape of time structure ($a(t^{\prime})$, see Eq.~(D.17))
including a uniform (rectangular) distribution.

\section{Signal charge distribution in the ideal case}

In this appendix, we present the signal charge distribution in the ideal case, i.e.  
in the absence of the charge spread in the GEM stack, the laser pulse-energy fluctuations and the electronic noise.
As in Appendix B, we assume that
the probability density function of $x \equiv g/\left< g \right>$,
the gas gain ($g$) normalized by its average ($\left< g \right>$), for single drift electrons
is given by
\begin{equation}
      P_1(x) = 
      A \cdot
      x^{\theta} \cdot
      \exp \displaystyle \Bigl( -(1 + \theta) \cdot x \Bigr)
\end{equation}
with
\begin{equation}
  A \equiv \frac{(1+\theta)^{1+\theta}}{\Gamma (1+\theta)} \;
\end{equation}
for the continuous random variable $x \; > \; 0$,
where
$\Gamma$ represents the gamma function defined in Appendix B (Eq.~(B.31)),
and $\theta$ ($>-1$) is a parameter determining the shape of distribution ($f = 1/(1+\theta)$).

The charge distribution for two drift electrons, scaled by $\left< g \right>$, is then given by
\begin{eqnarray}
  P_2(x) &=& A^2 \cdot \int_0^x dx_1 \int_0^x dx_2 \; (x_1 \cdot x_2)^\theta \cdot e^{-(1+\theta)\cdot (x_1+x_2)} \cdot \delta(x - x_1 - x_2) \\
  &=& A^2 \cdot e^{-(1+\theta)x} \cdot \int_0^x dx_1 \bigl( x_1 \cdot (x-x_1) \bigr)^\theta \\
  &=& A^2 \cdot e^{-(1+\theta)x} \cdot \int_0^1 (x \cdot t)^\theta \cdot  (x-x\cdot t)^\theta \cdot x \cdot dt \;\;\; {\rm with} \;\;  t \equiv \frac{x_1}{x} \\
  &=& A^2 \cdot e^{-(1+\theta)x} \cdot x^{(1+2\theta)} \cdot \int_0^1 t^\theta (1 - t)^\theta dt \\
  &=& A^2 \cdot B(1+\theta, 1+\theta) \cdot x^{1+\theta} \cdot x^\theta \cdot e^{-(1+\theta)x} 
\end{eqnarray}
where $B$ represents the beta function
%%%\begin{eqnarray}
%%%  B(\alpha, \beta) &\equiv& \int_0^1 t^{\alpha - 1} \cdot (1-t)^{\beta - 1} \; dt \\
%%%  &=& \frac{\Gamma (\alpha) \cdot \Gamma(\beta)}{\Gamma (\alpha + \beta)} \;.
%%%\end{eqnarray}
\begin{equation}
  B(\alpha, \beta) \equiv \int_0^1 t^{\alpha - 1} \cdot (1-t)^{\beta - 1} \; dt = \frac{\Gamma (\alpha) \cdot \Gamma(\beta)}{\Gamma (\alpha + \beta)} \;.
\end{equation}
The normalization constant is expressed as
\begin{eqnarray}
  A^2 \cdot B(1+\theta, 1+\theta) &=& \left( \frac{(1+\theta)^{1+\theta}}{\Gamma (1+\theta)} \right)^2 \cdot \frac{\left( \Gamma (1+\theta) \right)^2}{\Gamma(2(1+\theta))} \\
      &=& \frac{(1+\theta)^{2(1+\theta)}}{\Gamma(2(1+\theta))} \;.
\end{eqnarray}
Accordingly
\begin{equation}
  P_2 (x) = \frac{(1+\theta)^{2(1+\theta)}}{\Gamma(2(1+\theta))} \cdot x^{1+\theta} \cdot x^\theta \cdot e^{-(1+\theta)x} \;.
\end{equation}  
\begin{comment}
It should be noted that
\begin{equation}
  \left< x \right> = \int_0^\infty x \cdot P_2(x) \; dx = 2 \;. 
\end{equation}
\end{comment}

Similarly, the charge distribution for three drift electrons scaled by $\left< g \right>$ is given by,
with
\begin{equation}
  x^\prime \equiv x_1 + x_2,\;\; C \equiv A^2 \cdot B(1+\theta,1+\theta) = \frac{(1+\theta)^{2(1+\theta)}}{\Gamma(2(1+\theta))},\;\;{\rm and } \; t \equiv \frac{x_3}{x} \;, 
\end{equation}
\begin{comment}
\begin{eqnarray}
  x^\prime = x_1 + x_2,\;\; C &=& A^2 \cdot B(1+\theta,1+\theta),\;\;{\rm and } \; t = \frac{x_3}{x} \;, \\ 
                            &=& \frac{(1+\theta)^{2(1+\theta)}}{\Gamma(2(1+\theta))}                                         
\end{eqnarray}
\end{comment}
\begin{eqnarray}
  P_3(x) &=& \int_0^x dx_3 \int_0^x dx^\prime \; P_1(x_3) \cdot P_2(x^\prime) \cdot \delta (x - x^\prime - x_3) \\
  &=& A \cdot\ C \cdot e^{-(1+\theta)x} \cdot \int_0^x dx_3 \; \int_0^x dx^\prime \; x_3^\theta \cdot (x^\prime)^{1+\theta} \cdot (x^\prime)^\theta \cdot \delta (x - x^\prime - x_3) \\
  &=& A \cdot\ C \cdot e^{-(1+\theta)x} \cdot \int_0^x dx_3 \; x_3^\theta \cdot \left( x - x_3 \right)^{1+ 2\theta} \\
  &=& A \cdot\ C \cdot e^{-(1+\theta)x} \cdot \int_0^1 x \; dt \; (x\cdot t)^\theta \cdot (x - x\cdot t)^{1+2\theta} \\
%%%  &=& A \cdot\ C \cdot e^{-(1+\theta)x} \cdot \int_0^1 x \; dt \; (x\cdot t)^\theta \cdot (x - x\cdot t)^{1+2\theta} \;\;\; {\rm with} \;\;  t \equiv \frac{x_3}{x}\\
  &=& A \cdot\ C \cdot e^{-(1+\theta)x} \cdot x^\theta \cdot x^{2(1+\theta)} \cdot \int_0^1 t^\theta \cdot (1-t)^{1+2\theta} \; dt \\
  &=& A \cdot\ C \cdot B(1+\theta, 2(1+\theta)) \cdot x^{2(1+\theta)} \cdot x^\theta \cdot e^{-(1+\theta)x} \;
\end{eqnarray}
where
\begin{equation}
  B(1+\theta, 2(1+\theta)) = \frac{\Gamma(1+\theta) \cdot \Gamma(2(1+\theta))}{\Gamma(3(1+\theta))} \;.
\end{equation}
The normalization constant is then
\begin{eqnarray}
  A \cdot C \cdot B(1+\theta, 2(1+\theta)) &=& A^3 \cdot B(1+\theta,1+\theta) \cdot B(1+\theta, 2(1+\theta)) \\
   &=& \frac{(1+\theta)^{1+\theta}}{\Gamma(1+\theta)} \cdot \frac{(1+\theta)^{2(1+\theta)}}{\Gamma(2(1+\theta))}
                             \cdot  \frac{\Gamma(1+\theta) \cdot \Gamma(2(1+\theta))}{\Gamma(3(1+\theta))} \\
   &=& \frac{(1+\theta)^{3(1+\theta)}}{\Gamma(3(1+\theta))} \;.
\end{eqnarray}
Therefore,
\begin{equation}
P_3 (x) = \frac{(1+\theta)^{3(1+\theta)}}{\Gamma(3(1+\theta))} \cdot x^{2(1+\theta)} \cdot x^\theta \cdot e^{-(1+\theta)x} \;.
\end{equation}

Since each increment of the number of drift electrons adds $1+\theta$ to the power of $x$,
with the exponential dependence preserved, the probability density function of the n-fold self-convolution
of a gamma distribution has the following $x$-dependence:
\begin{equation}
  P_n(x) \propto x^{(n-1)\cdot (1+\theta)} \cdot x^\theta \cdot e^{-(1+\theta)x} \;.
\end{equation}
The normalization constant is given by
\begin{eqnarray}
  A^n \cdot \prod_{k=1}^{n-1} B(1+\theta, k(1+\theta)) &=& A^n \cdot \prod_{k=1}^{n-1} \frac{\Gamma(1+\theta) \cdot \Gamma(k(1+\theta))}{\Gamma((k+1) (1+\theta))} \\
  &=& A^n \cdot \frac{\left( \Gamma (1+\theta) \right)^n}{\Gamma(n(1+\theta))} \\
  &=& \frac{(1+\theta)^{n(1+\theta)}}{\Gamma(n(1+\theta))} \;.
\end{eqnarray}
It can be obtained by the integration as well: with $\xi = (1+\theta)\cdot x$,
\begin{eqnarray}
  \int_0^\infty x^{(n-1)\cdot (1+\theta)} \cdot x^\theta \cdot e^{-(1+\theta)x} \; dx &=&  \int_0^\infty \left( \frac{\xi}{1+\theta} \right)^{n\cdot \theta + n -1} \cdot e^{-\xi}
           \cdot \frac{d\xi}{1+\theta} \\
           &=& \frac{1}{(1+\theta)^{n(1+\theta)}} \cdot \int_0^\infty \xi^{n\cdot \theta + n - 1} \cdot e^{-\xi} \; d\xi \\
           &=& \frac{\Gamma(n(1+\theta))}{(1+\theta)^{n(1+\theta)}} \;.
\end{eqnarray}
Accordingly, the n-fold self-convolution of a Polya (gamma) distribution is expressed by 
\begin{equation}
  P_n(x) = \frac{(1+\theta)^{n(1+\theta)}}{\Gamma(n(1+\theta))} \cdot x^{(n-1)(1+\theta)} \cdot x^\theta \cdot e^{-(1+\theta)x} \;.
\end{equation}

The average of $x$ is given by (with $\xi = (1+\theta)\cdot x$)
\begin{eqnarray}
  \langle x \rangle &=& \int_0^\infty x \cdot P_n (x) \; dx \\
  &=& \frac{(1+\theta)^{n(1+\theta)}}{\Gamma(n(1+\theta))} \cdot \int_0^\infty x^{n(1+\theta)} \cdot e^{-(1+\theta)x} \; dx \\
  &=& \frac{(1+\theta)^{n(1+\theta)}}{\Gamma(n(1+\theta))} \cdot \int_0^\infty \left( \frac{\xi}{1+\theta} \right)^{n(1+\theta)} \cdot
                          e^{-\xi} \cdot \frac{d\xi}{1+\theta} \\
                          &=& \frac{1}{(1+\theta)\cdot \Gamma(n(1+\theta))} \cdot \int_0^\infty \xi^{n(1+\theta)} \cdot e^{-\xi} \; d\xi \\
                          &=& \frac{\Gamma(n(1+\theta) + 1)}{(1+\theta) \cdot \Gamma(n(1+\theta))} \\
                          &=& \frac{n \cdot (1+\theta) \cdot \Gamma(n(1+\theta))}{(1+\theta) \cdot \Gamma(n(1+\theta))} \\
                          &=& n \;.
\end{eqnarray}
It should be noted that $\Gamma (z+1) = z \cdot \Gamma (z)$.

In the absence of the charge spread in the GEM stack, the number of drift electrons detected by a pad row
has a Poisson distribution with an average $\left< n \right>$.
The signal charge then follows a compound Poisson-Polya (gamma) distribution: 
\begin{equation}
  P(x) = e^{-\left< n \right>} \cdot \left( \delta (x) + \sum_{n=1}^\infty \frac{\left< n \right>^n}{\Gamma(n+1)} \cdot P_n(x) \right) \;.
\end{equation}
The above formula is identical to that assumed in Ref. \cite{arnaud}.

The gamma distribution, i.e. a Polya distribution for a continuous domain ($x \in \mathbb{R}^+$),
allows us a simple analytical treatment.
See Section 6.4 for the (possible) justification of the Polya (gamma) distribution for the avalanche fluctuations in GEM stacks.

\section{Avalanche fluctuations given by the Legler's model and the gain stability}

In this appendix, we discuss the minimum possible value of avalanche fluctuations
and the gas-gain stability for a perfect parallel-plate geometry
based on the Legler's model~\cite{Legler0, Legler}.
The simple model given by Legler supposes that the electric field is not too high and the kinetic energy of electrons
before an ionizing collision is not too large so that the two electrons just after the ionization have a small amount of kinetic energy
compared to the ionization potential of the gas.
It should be kept in mind that
this condition is assumed to be fulfilled in the following argument
even under high reduced electric fields ($E/N$)
with $N$, the density of gas atoms and/or molecules.
In addition, only impact ionizations (free from space-charge effects) are assumed
even for large gas gains.

The asymptotic relative variance of the size of avalanches initiated by an electron,
in the limit of large average size, is given by~\cite{alkhazov}
\begin{equation}
f_0 = \frac{(2 \cdot b - 1)^2}{4 \cdot b - 2 \cdot b^2 -1}\;, \quad b \in \left[ \; 1/2, \; 1 \; \right]
\end{equation}
for parallel plate geometries (with a uniform electric field $E$ and the first Townsend coefficient $\alpha$), with
\begin{eqnarray}
  b &\equiv& \exp( - \chi^\prime ) \\
  \chi^\prime &\equiv& \chi \cdot \frac{U_0}{U_{\rm i}} \equiv \alpha \cdot x_0
\end{eqnarray}
where  $U_{\rm i}$ is the the ionization potential of the gas atom or molecule,
$U_0$ is a model parameter, $x_0 \equiv U_0 / E$ denotes the relaxation distance, and $\chi$ is defined by
\begin{equation}
  \chi \equiv \alpha \cdot \frac{U_{\rm i}}{E} \;.
\end{equation}
Another parameter often used ($H$) is the reciprocal of $\chi$.

It should be noted that the Townsend coefficient for electrons in equilibrium with the electric field ($a$) is given by
\begin{equation}
  a = \frac{\alpha}{2 \cdot b - 1}\;.
\end{equation}
The lower limit of $b$ is determined from $a \geq 0$, and the relation
\begin{equation}
  b \equiv \exp \left( -\alpha \cdot \frac{U_0}{E} \right) \geq \frac{1}{2}
\end{equation}
restricts the value of the model parameter $U_0$:
\begin{equation}
\alpha \cdot \frac{U_0}{E} = \chi^\prime \leq \ln{2} \;.
\end{equation}
It should also be noted that $f_0$ is an increasing function of $b$. 
When $b$ is maximum (unity) $f_0$ takes the maximum value (unity) with the relaxation distance vanished,
and a geometric (exponential) distribution represents the (asymptotic normalized) avalanche fluctuations
(see, for example, Refs.~\cite{wijsman,legler-1955}).
Hence, $f_0$ is minimum when the ionization efficiency ($\eta \equiv \alpha / E$) is maximum.\footnote{
  The symbol $\eta$ here is the notation used in Ref.~\cite{kruithof} for the {\it ionization efficiency\/}
  defined as the number of ion pairs created by an electron traveling through a potential difference of one volt,
  and is not the attachment coefficient as conventionally used.
  }

Let $d$ be the the distance between the anode and the cathode.
When $d$ is large enough the average avalanche size ($\left< G \right>$) is given by
\begin{eqnarray}
  \left< G \right> &\sim& \exp (\alpha \cdot d) \\
  &\sim& \exp\left(\alpha \cdot \frac{V}{E}\right) \\
  &\sim& \exp(\eta \cdot V) 
\end{eqnarray}
where $V$ denotes the voltage applied between the anode and the cathode ($V = E \cdot d$).
For a fixed value of $V$, therefore, 
$\left< G \right>$ is maximum and $f_0$ is minimum when $\eta$ is maximum ($\eta^\ast$).

Now, $\eta$ can be expressed as, with a function $F(E/N)$,
\begin{eqnarray}
  \eta \left( \frac{E}{N} \right) &\equiv& \frac{\alpha}{E} \\
  &=& \frac{N}{E} \cdot F\left( \frac{E}{N} \right) \\
  &=& \frac{d\cdot N}{V} \cdot F\left( \frac{V}{d \cdot N} \right) \\
  &=& \eta \left( d \cdot N \right)
\end{eqnarray}
for a fixed $V$.
Let $d^\ast$ be the electrode distance that gives $\eta^\ast$ for a given $N$.
Then, if $d$ is set to $d^\ast$ the average gas gain ($\left< G \right>$) is maximum 
and is stable against the variations of the electrode distance and/or the gas density,
and the relative variance ($f_0$) is minimum. 
%%%%%The actual relative variance $f$ may be a little smaller than $f_0$ because the latter is the
%%%%%asymptotic value of $f$ in the limit of $d \to \infty$ and $f$ is an increasing function of $d$.

It should be noted, however, that $\eta^\ast$ is usually realized at a rather large value of $E/N$,
that could make the chamber prone to discharge.
In addition, a rather narrow gap ($d^\ast$) may be required to keep the average gas gain moderate.
See, for example, Fig.~3 in Ref.~\cite{alkhazov} for the ionization efficiencies ($\eta$) of
argon and methane as function of $E/N$. 
The figure tells us that the electric fields corresponding to $\eta^\ast$ are
$\gsim$ 100 kV/cm$\cdot$atm and $\gsim$ 200 kV/cm$\cdot$atm, respectively for pure argon and pure methane.
Actually shown in the figure are the behaviors of $\chi (E/N) \equiv U_{\rm i} \cdot \eta$,
that is the ratio of the energy spent by a drifting electron for successive ionizations
to the total energy supplied to the electron by the electric field.
It is therefore naively expected that $\left< G \right>$ is maximum while $f_0$ is minimum when $\eta$ is maximum.

It may be worth noting that $f_0 = 0$ when $b = 1/2$, the minimum allowed value corresponding to $\chi^\prime = \ln{2}$
(see Eqs.~(F.6) and (F.7)).
In this case,
\begin{equation}
   G = \exp \left( \chi^\prime \cdot \frac{d}{x_0} \right) = 2^{\frac{d}{x_0}}  
\end{equation}
without fluctuations,
corresponding to Model 2 in Ref.~\cite{alkhazov} with $a = \infty$
(a 100\% instantaneous ionizing collision after every step $x_0$ for each of electrons in an avalanche).  

In reality, however, the minimum value of $f_0$ is determined by the maximum of $\eta$ ($\eta^\ast$) as mentioned above.
For pure argon, for example, $\eta^\ast \sim$ 0.0221 at $E \sim$ 150 kV/cm at atmospheric pressure~\cite{kruithof}.
Then, assuming $U_0 = U_{\rm i} = 15.76$ eV~\cite{Gstir}, $\chi^\prime = \chi = \eta^\ast \cdot U_{\rm i} \sim 0.349$ and $b \sim 0.706$,
and the minimum of $f_0 \sim 0.21$.\footnote{
  For pure neon $\eta^\ast \sim 0.0149$ at $E \sim$ 65 kV/cm$\cdot$atm~\cite{kruithof}, giving the minimum of $f_0 \sim 0.24$
  with $U_0 = U_{\rm i} = 21.56$ eV~\cite{Saloman}.
  }

For an {\it ideal\/} Micromegas with an abrupt transition of the electric field strength at the mesh plane, 
the energy resolution ($R$) for the $^{55}$Fe main (K$_\alpha$) peak would be given by
\begin{equation}
R^2 = \frac{F + f_0}{N_{\rm e}}
\end{equation}
where $F$ is the Fano factor~\cite{Fano} and $N_{\rm e}$ is the average number of electrons created by 5.89 keV photon quanta.
If we assume $F$ = 0.17~\cite{iaea-tecdoc-799} and the $W$-value of 26.4 eV for pure argon~\cite{icru79}, then $N_{\rm e}$ = 223 and
the resolution (FWHM) is expected to be 9.7\%.
The operation at the optimum point has already been investigated with a small gap Micromegas, despite the difficulty mentioned above~\cite{attie}. 
For a more moderate (realistic) value of $E$ = 30 kV/cm$\cdot$atm
$f_0 \sim 0.50$ for argon with $U_0 = U_{\rm i} = 15.76$ eV whereas $f_0 \sim 0.28$ for neon with $U_0 = U_{\rm i} = 21.56$ eV.

It is a common practice to add a hydrocarbon gas to a rare gas as a quencher for UV photons created in electron avalanches.
The hydrocarbon quencher (isobutane in our case) cools down electrons in the avalanche by its large inelastic cross sections to relatively low-energy electrons.
On the other hand, it helps avalanche growth when its ionization potential is lower than the potential energies of the excited states of the rare gas atom
by the Penning effect~\cite{Penning}.
The net effect of adding a hydrocarbon quencher can be the increase of gas gain when its concentration is small enough so that
its role as a Penning agent is more efficient than as a moderator.
A small amount of isobutane in our gas mixture drastically decreases the operating high voltages of GEMs
(see, for example,  Figs.~6 and 7 in Ref.~\cite{cf4paper}).
In this case, the ratio of the energy spent for ionizations to that supplied to the electrons by the electric field increases, thereby
decreasing the value of $f_0$.
In addition, the number of ion pairs created by 5.89 keV photons ($N_{\rm e}$) is expected to increase with a reduced $W$-value~\cite{bronic}.  
Hence, the energy resolution for the main peak of $^{55}$Fe given by Eq.~(F.16) is expected to improve
as far as the photoelectric effect due to UV photons is well suppressed.

\vspace{-2.5mm}

\section{Avalanche fluctuations for a double GEM stack}
\setcounter{figure}{0}
\renewcommand{\thesubsection}{\Alph{section}.\arabic{subsection}}
\setcounter{section}{7}
\setcounter{subsection}{0}
In this appendix, we present an approximate estimate for the relative variance
of the response of a double GEM stack to single electrons.
We assume a perfect stack of GEMs with identical holes in the flat and uniform foil,
and constant gaps between the GEMs and between the second GEM and a readout plane.
The estimation can readily be extended for a stack of three or more GEMs.

\subsection{Response of a single GEM to multiple electrons}

\vspace{5mm}

Let $n$ and $\sigma_{\rm n}$ be the number of incoming electrons and their fluctuations.
The electrons are gas-amplified through the following processes:
\begin{enumerate}
\item Collection into a GEM holes with an efficiency $p_{\rm 1}$
\item Gas multiplication with a gain $M$ (absolute gain)
\item Extraction to the transfer or induction gap with an efficiency $p_{\rm 2}$.
\end{enumerate}
We assume binomial statistics for both the collection and extraction processes.
In addition, independent binomial selections are assumed for the extraction process and the subsequent collection process
in the case of double GEMs considered later.
See Fig.~G.1 for the notation to be used.

%%%%%%%%%%%%%%%%%%%%%%%%%%%%%%%%%%%%%%%%%%%%%%%%%%%%%%%%%%%%%%%%%%%%%%%%%%%%%%
\begin{figure}[htbp]
\begin{center}
{    
\includegraphics*[scale=0.37]{\figdir/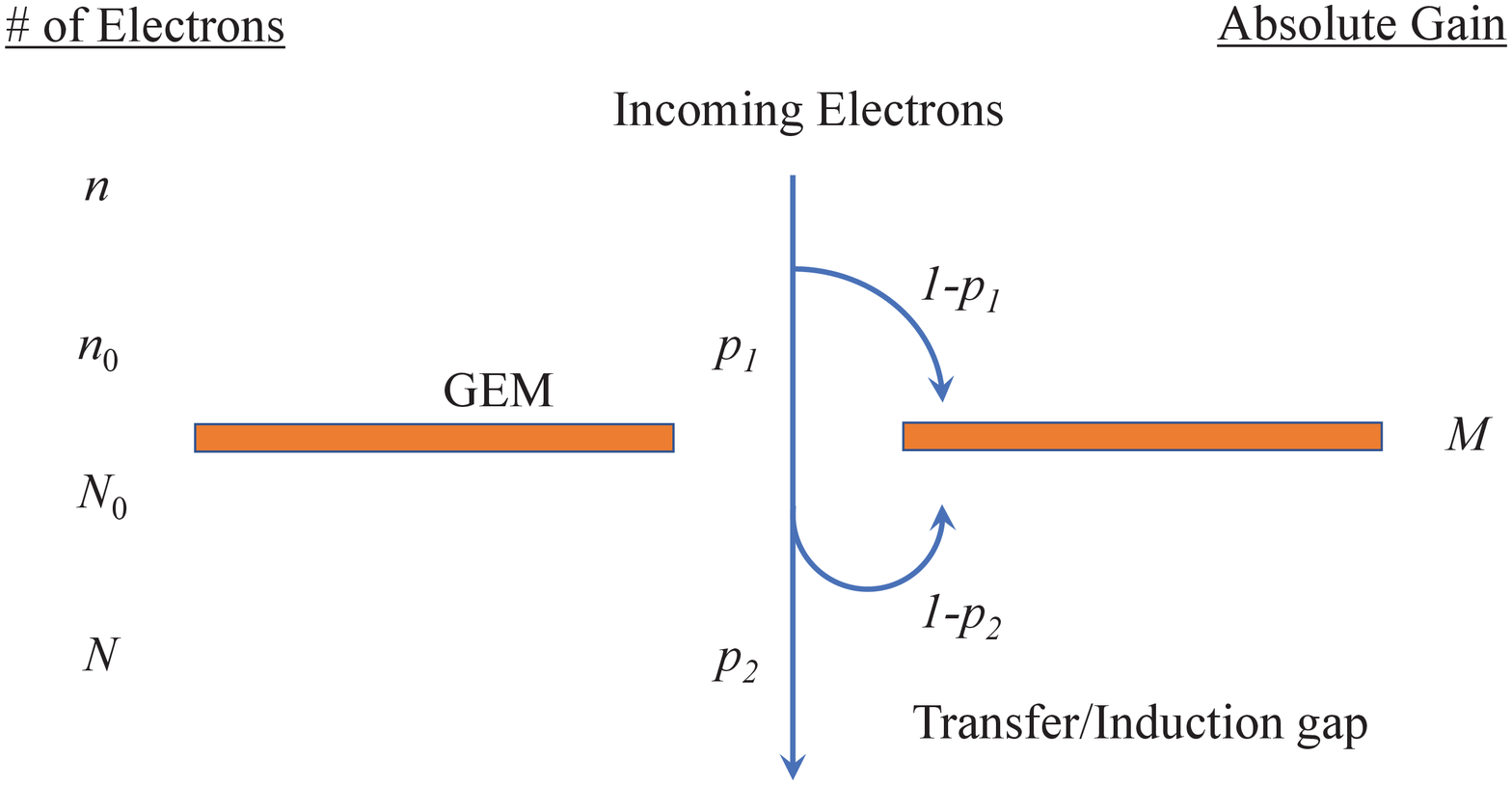}
}
\caption[fig1]{\label{fig1}
 Notation.
}
\end{center}
\end{figure}
%%%%%%%%%%%%%%%%%%%%%%%%%%%%%%%%%%%%%%%%%%%%%%%%%%%%%%%%%%%%%%%%%%%%%%%%%%%%%%

The average number of electrons entering the GEM holes ($n_0$) is 
\begin{equation}
  \left< n_0 \right>  = p_1 \left< n \right> \;. 
\end{equation}
The variance of $n_0$ is given by 
\begin{eqnarray}
  \left< \left( n_0 - \left< n_0 \right> \right)^2 \right> &=& \left< \left( n_0 - p_1 \cdot \left< n \right> \right)^2 \right> \\  
  &=& \left< \Bigl( n_0 - p_1 \cdot n + p_1 \cdot \left( n - \left< n \right> \right) \Bigr)^2 \right> \\
  &=& \left< \left( n_0 - p_1 \cdot n \right)^2 \right> + p_1^2 \cdot \left< \left( n - \left< n \right> \right)^2 \right> \\
  &=& \Bigl< n \cdot p_1 \cdot \left( 1 - p_1 \right) \Bigr> + p_1^2 \cdot \left< \left( n - \left< n \right> \right)^2 \right> \\
  &=& \left< n \right> \cdot p_1 \cdot \left( 1 - p_1 \right) + p_1^2 \cdot \sigma_{\rm n}^2\; . 
\end{eqnarray}
Accordingly, the relative variance of $n_0$ is given by
\begin{eqnarray}
  RVar\,(n_0) &\equiv& \frac{\sigma_{\rm n_0}^2}{\left< n_0 \right>^2} \\
  &=& \frac{ \left< n \right> \cdot p_1 \cdot \left( 1 - p_1 \right) + p_1^2 \cdot \sigma_{\rm n}^2}{p_1^2 \cdot \left< n \right>^2} \\
  &=& \frac{1-p_1}{p_1} \cdot \frac{1}{\left< n \right>} + \frac{\sigma_{\rm n}^2}{\left< n \right>^2} \\
  &=& RVar\,(n) + \frac{1}{\left< n \right>} \cdot \frac{1-p_1}{p_1}  
\end{eqnarray}
with
\begin{equation}
RVar\,(n) \equiv \frac{\sigma_{\rm n}^2}{\left< n \right>^2} \;.
\end{equation}
The relative variance of $n_0$ is the sum of that of the number of incoming electrons and
$(1-p_1)/p_1$ divided by the average number of incoming electrons ($\left< n \right>$).

The number of gas-amplified electrons ($N_0$) is then
\begin{equation}
  N_0 = \sum_{i=1}^{n_0} M_i
\end{equation}
where $M_i$ is the absolute gas gain of the GEM for the $i$-th incoming electron.
The average of $N_0$ is 
\begin{equation}
  \left< N_0 \right> = \left< n_0 \right> \left< M \right> = p_1 \cdot \left< n \right> \cdot \left< M \right>
\end{equation}
and its variance is given by 
\begin{eqnarray}
  \left< \left( N_0 - \left< N_0 \right> \right)^2 \right> &=& \left< \left( \sum_{i=1}^{n_0} M_i - \left< n_0\right>  \cdot \left< M \right> \right)^2 \right> \\
  &=& \left< \left( \sum_{i=1}^{n_0} \left( M_i - \left< M \right> \right) + \left( n_0 - \left< n_0 \right> \right) \cdot \left< M \right> \right)^2 \right> \\
  &=& \bigg\langle n_0 \cdot \left< \left( M - \left< M \right> \right)^2 \right> \bigg\rangle + \left< \left( n_0 - \left< n_0 \right> \right)^2 \right> \cdot \left< M \right>^2 \\
  &=& \left< n_0 \right> \cdot \sigma_{\rm M}^2 + \left< M \right>^2 \cdot \sigma_{\rm n_0}^2
\end{eqnarray}
where $M_i$ and $M_j$ ($j \neq i$) are assumed to be independent of each other.
Accordingly, the relative variance is given by
\begin{eqnarray}
  RVar\,(N_0) &\equiv& \frac{\sigma_{\rm N_0}^2}{\left< N_0 \right>^2} \\
  &=& \frac{\left< n_0 \right> \cdot \sigma_{\rm M}^2 + \left< M \right>^2 \cdot \sigma_{\rm n_0}^2}{\left< n_0 \right>^2 \cdot \left< M \right>^2} \\
  &=& \frac{\sigma_{\rm n_0}^2}{\left< n_0 \right>^2} + \frac{1}{\left< n_0 \right>} \cdot \frac{\sigma_{\rm M}^2}{\left< M \right>^2} \\
  &=& RVar\,(n_0) + \frac{1}{\left< n_0 \right>} \cdot RVar\,(M) \;.
\end{eqnarray}

The average number of extracted electrons ($N$) is therefore
\begin{equation}
  \left< N \right> = p_2 \cdot \left< N_0 \right> = p_1 \cdot p_2 \cdot \left< M \right> \cdot \left< n \right> 
\end{equation}
and its variance is given by
\begin{eqnarray}
  \left< \left( N - \left< N \right> \right)^2 \right> &=& \left< \left( N - p_2 \cdot \left< N_0 \right> \right)^2 \right> \\
  &=& \left< \Bigl( N - p_2 \cdot N_0 + p_2 \cdot \left( N_0 - \left< N_0 \right> \right) \Bigr)^2 \right> \\
  &=& \left< \left( N - p_2 \cdot N_0 \right)^2 \right> + p_2^2 \cdot \left< \left( N_0 - \left< N_0 \right> \right)^2 \right> \\
  &=& \left< N_0 \right> \cdot p_2 \cdot \left( 1 - p_2 \right) + p_2^2 \cdot \sigma_{\rm N_0}^2 \;.
\end{eqnarray}
The relative variance of $N$ is then given by
\begin{eqnarray}
  RVar\,(N) &\equiv& \frac{\sigma_{\rm N}^2}{\left< N \right>^2} \\
  &=& \frac{ \left< N_0 \right> \cdot p_2 \cdot \left( 1 - p_2 \right) + p_2^2 \cdot \sigma_{\rm N_0}^2}{p_2^2 \cdot \left< N_0 \right>^2} \\
  &=& \frac{1}{\left< N_0 \right>} \cdot \frac{1 - p_2}{p_2} + RVar\,( N_0) \\
  &=& RVar\,(n_0) + \frac{1}{\left< n_0 \right>} \cdot RVar\,(M) + \frac{1}{\left< N_0 \right>} \cdot \frac{1 - p_2}{p_2} \\
  &=& RVar\,(n) + \frac{1}{\left< n\right>} \cdot \frac{1 - p_1}{p_1} + \frac{1}{p_1 \cdot \left< n \right>} \cdot RVar\,(M)
                     + \frac{1}{p_1 \cdot \left< n \right> \cdot \left< M \right>} \cdot \frac{1 - p_2}{p_2} \;. 
\end{eqnarray}

\subsection{In the case of double GEM stack}

\vspace{5mm}

For the first stage, $n = 1$, $\sigma_{\rm n} = 0$, $p_1 = 1$, and
\begin{eqnarray}
  \left< N_1 \right> &=& p_2 \cdot \left< M \right> \\
  RVar\,(N_1) &=& RVar\,(M) + \frac{1}{\left< M \right>} \cdot \frac{1- p_2}{p_2} 
\end{eqnarray}
where $N_1$ denotes the number of electrons entering the transfer gap.
Including the second stage with $p_1^{\prime}$, $p_2^{\prime}$, and $M^{\prime}$,
the average of the total number of electrons ($N_2$) is 
\begin{equation}
\left< N_2 \right> = p_2 \cdot p_1^{\prime} \cdot p_2^{\prime} \cdot \left< M \right> \cdot \left< M^{\prime} \right>
\end{equation}
and its relative variance is given by
\begin{eqnarray}
  RVar\left(N_2\right) &=& RVar\left( N_1\right) + \frac{1}{\left< N_1\right>} \cdot \frac{1-p_1^{\prime}}{p_1^{\prime}}
  + \frac{1}{p_1^{\prime} \cdot \left< N_1 \right>} \cdot RVar\,(M^{\prime}) \nonumber \\
  & & \hspace{75mm}
    + \frac{1}{p_1^{\prime} \cdot \left< N_1 \right> \cdot \left< M^{\prime} \right>} \cdot \frac{1-p_2^{\prime}}{p_2^{\prime}} \\
    &=& RVar\,(M) + \frac{1}{\left< M \right>} \cdot \frac{1-p_2}{p_2} 
    + \frac{1}{p_2 \cdot \left< M \right>} \cdot \frac{1-p_1^{\prime}}{p_1^{\prime}} \nonumber \\
  & & \hspace{25mm}
    + \frac{1}{p2 \cdot p_1^{\prime} \cdot \left< M \right>} \cdot RVar\,(M^{\prime}) 
    + \frac{1}{p_2 \cdot p_1^{\prime} \cdot \left< M \right> \cdot \left< M^{\prime} \right>} \cdot \frac{1 - p_2^{\prime}}{p_2^{\prime}} \;.   
\end{eqnarray}
The sum of the second and the third terms in Eq.~(G.36) can be expressed as
\begin{equation}
\frac{1}{\left< M \right>} \cdot \frac{1-p_2}{p_2}
         + \frac{1}{p_2 \cdot \left< M \right>} \cdot \frac{1-p_1^{\prime}}{p_1^{\prime}}
  = \frac{1}{\left< M \right>} \cdot \frac{1 - p_2 \cdot p_1^{\prime}}{p_2 \cdot p_1^{\prime}}  
\end{equation}
since the two consecutive binomial selections, one at the exit of the first GEM
and the other at the entrance to the second GEM are assumed to be independent.\footnote{
  Strictly speaking, the sum of the two terms depends on the relative alignment of the GEM holes in the upper and lower GEMs,
  as well as on the diffusion of amplified electrons in the transfer gap.
  The holes are usually neither aligned nor completely misaligned. 
  Therefore, the assumption is justified when the diffusion in the transfer gap is large
  compared to the GEM hole pitch.
  It is to be noted that this is not always the case; the diffusion is about 265 $\mu$m in standard deviation
  whereas the GEM hole pitch is 140 $\mu$m for our GEM stack. 
  Fortunately, the contribution of the transfer efficiency fluctuations is suppressed by the effective gain of the first GEM ($p_2 \cdot \left< M \right>$).
}

\subsection{Evaluation for a typical double GEM stack}

Let us assume the absolute gas gains of the two GEM foils are identical just for simplicity
($\left< M^{\prime} \right> = \left< M \right>$ and $RVar\,(M^{\prime}) = RVar\,(M)$).
Then the relative variance of a double-GEM stack is given by
\begin{equation}
  RVar\,(N_2) = RVar\,(M) + \frac{1}{\left< M \right>}\frac{1-p}{p} + \frac{RVar\,(M)}{p \cdot \left< M \right>}
  + \frac{1}{p \cdot \left< M \right>^2} \cdot \frac{1-p_2^{\prime}}{p_2^{\prime}}
\end{equation}
with
\begin{equation}
  p \equiv p_2 \cdot p_1^{\prime} \;.
\end{equation}
Let us further assume typical values for the extraction and collection efficiencies~\cite{sauli2006}:
$p_2 = p_2^{\prime} = 40$\% and $p_1^{\prime} = 80$\%,  
and 170 for the absolute gas gain ($\left< M \right>$).
The assumed total effective gain is then close to that of our GEM stack ($\sim$3700).

If we put a typical value of 2/3 for $RVar\,(M)$ 
Eq.~(G.38) yields $\sim$0.69 for the total relative variance under the conditions mentioned above. 
The relative variance is dominated by that of the first stage and the contribution of the following stage(s)
is small, as naively expected,
as far as the average effective gain of the first stage ($p_2 \cdot \left< M \right>$) is large enough.

\end{document}